\documentstyle[12pt]{article}

\input{tcilatex}

\begin{document}

\begin{titlepage}
\vspace{1.5cm} 
\begin{center}
{\Large\bf  ON KMS CONDITION FOR THE CRITICAL ISING MODEL}\footnote{Work partially supported  by CNPq-BRASIL}\\
\vspace{1cm}
{\bf   A. Lima-Santos}
\hspace{.4cm}
          \bf   and 
\hspace{.4cm}
 { \bf        Wagner Utiel}\\
\vspace{1cm}
\begin{center}
Departamento de F\'{\i}sica\\
 Universidade Federal de S\~ao Carlos\\ 
  13569-905 S\~ao Carlos, Brasil\\ 
\end{center}
\vspace{2.0cm}
     \begin{abstract}

 Using the KMS condition and exchange algebras we discuss the monodromy 
and modular properties of two-point KMS states of the Critical Ising Model. 

 \end{abstract}
 \end{center}
\end{titlepage}
\newpage 

In the chiral field theory one can find two different notions of modularity
present in literature. The one is related to transformation properties of
the characters of the theory under of the action of the modular group\cite
{Verlinde}, and the other kind of modularity stems from the Tomita-Takesaki
theory and describes the symmetry properties of temperature states\cite{BRA}%
. It should be a interesting question whether this common occurence of the
terminology "modularity" are related.

In this letter we tackle this problem by exploiting the
Kubo-Martin-Schwinger (KMS) condition \cite{KMS} in the context of conformal
field theory \cite{BMT} and the exchange algebra of the chiral conformal
field theories \cite{R} to compute the chiral KMS states for the critical
Ising model. We compare the KMS boundary condition and the action of the
center of the conformal group with the derived monodromy properties of the
chiral KMS states. The analysis of the short-distance singularities of their
normalized combinations leads to modular properties of the characters of the
simplest minimal model.

If one denotes by ${\cal A}$ a chiral field algebra and by $\alpha _{t}$ ( $%
t\in R$ ) the automorphism of ${\cal A}$ inducing the time translation in
the considered heat bath, the KMS condition reads as: The state $\omega
_{\beta }$ on ${\cal A}$ satisfies the KMS condition at inverse temperature $%
\beta >0$ if and only if for every pair of operators $A,B$ $\in $ ${\cal A}$
there exists an analytic function $F$ in the strip $S_{\beta }=\{Z\in C\mid
0<{\rm Im}$ $Z<\beta \}$ with continuous boundary values at $\func{Im}Z=0$
and $\limfunc{Im}Z=i\beta $, given respectively by $F(t)=\omega _{\beta
}(A\alpha _{t}(B))$ and $F(t+i\beta )=\omega _{\beta }(\alpha _{t}(B)A)$.

Due to energy positivity, the function $F(t)$ is holomorphic in the strip $%
0<ImZ<\beta $ and satisfies the KMS boundary condition 
\begin{equation}
\omega _{\beta }(A\,\alpha _{t+i\beta }(B))=\omega _{\beta }(\alpha
_{t}(B)\,A)  \label{eq1.1}
\end{equation}
where the time-evolution automorphism $\alpha _{t}$ is defined by $\alpha
_{t}(A)=e^{iHt}A\,e^{-iHt}$ .

If $e^{-\beta H}$ is a trace-class operator, in the sense that the chiral
partition function $Z=Tr(e^{-\beta H})$ is well defined , then $\omega
_{\beta }(A)$ is given by the density matrix 
\begin{equation}
\omega _{\beta }(A)\doteq \ <A>_{\beta }=\frac{1}{Z(\beta )}Tr(e^{-\beta
H}A).  \label{eq1.2}
\end{equation}
$H$ is bounded below and the KMS condition (\ref{eq1.1}) reflects stability
and cyclic property of the trace which is carried over the eigenstates of $%
sH. $ It yields to the Ward identities on the torus \cite{Bernard} and in
simple field theories, like free fermions, it uniquely determines the
vacuum. An application of this condition to the Hawking radiation of the
black holes has been described by Haag, Narnhofer and Stein \cite{Haag}.

For conformal field models, $H$ will be replaced by the conformal
Hamiltonian $L_0\ +\stackrel{\_}{L}_0$ and we will be dealing with
representations for which the trace (\ref{eq1.2}) does exist.

In principle the relation (\ref{eq1.1}) can be used for the computation of
the possible finite-temperature states \cite{BRA}. In the light-cone field
theory we have a simple algebraic relation between the operators $A\alpha
_t(B)$ and $\alpha _t(B)A$, the exchange algebra for chiral fields \cite{R}.

The compact picture of the stress-energy tensor, say $T(x_{+})\rightarrow 
{\cal T}(\alpha )\quad $($x_{+}=2i\tan (\pi \alpha )$), has the following
Laurent expansion 
\begin{equation}
{\cal T}(\alpha )=\sum_{n\in Z}\stackrel{\sim }{L}_ne^{-2i\pi n\alpha
},\qquad \stackrel{\sim }{L}_n=L_n-\frac c{24}\delta _{n,0.}  \label{eq1.3}
\end{equation}

In order to compute the KMS states one can use the factorized form of the
local conformal field $\Phi (\alpha ,\stackrel{\_}{\alpha })$ in terms of
interpolating fields. Here this is done by considering the $2$-point
function $<\Phi (\alpha ,\stackrel{\_}{\alpha })\Phi (0,0)>_{\beta }$ as a
product of their chiral parts $<A(\alpha )A(0)>_{\beta }$ and $<A(\stackrel{%
\_}{\alpha })A(0)>_{\beta }$where , for instance 
\begin{equation}
<A(\alpha )A(0)>_{\beta }=\frac{Tr(e^{-\beta \stackrel{\sim }{L}%
_{0}}A(\alpha )A(0))}{\stackrel{\sim }{Z}(\beta )},  \label{eq1.4}
\end{equation}
where $\stackrel{\sim }{Z}(\beta )$ is the chiral partition function 
\begin{equation}
\stackrel{\sim }{Z}(\beta )=Tr(e^{-\beta \stackrel{\sim }{L}_{0}}),\qquad 
\stackrel{\sim }{L}_{0}=L_{0}-\frac{c}{24}\text{ .}  \label{eq1.5}
\end{equation}
and similar expressions for the other chiral part.

Introducing orthogonal projectors we can write

\begin{eqnarray}
\stackrel{\sim }{Z}(\beta )\sum_{\lambda ,\lambda ^{\prime }} &<&P_{\lambda
}\,A(\alpha )\,P_{\lambda ^{\prime }}\,A(0)\,P_{\lambda }>=\sum_{\lambda
,\lambda ^{\prime }}Tr(e^{2i\pi \tau \stackrel{\sim }{L_{0}}}P_{\lambda
}\,A(\alpha )\,P_{\lambda ^{\prime }}\,A(0)\,)  \nonumber \\
\ &\doteq &\sum_{\lambda ,\lambda ^{\prime }}{\cal F}_{\lambda \lambda
^{\prime }}(\alpha \mid \tau )  \label{eq1.6}
\end{eqnarray}
where we have introduced a new variable $\tau $ through the relation $2i\pi
\tau =-\beta .$

In the compact picture the analytically continued time evolution
automorphism of the field $A(\alpha )$ is given by $\alpha _{i\beta
}(A(\alpha ))=A(\alpha +\tau )$ . Therefore we can write the KMS condition
as 
\begin{equation}
\sum_{\lambda ,\lambda ^{\prime }}Tr(e^{2i\pi \tau \stackrel{\sim }{L_{0}}%
}P_{\lambda }\,A(\alpha +\tau )\,P_{\lambda ^{\prime
}}\,A(0)\,)=\sum_{\lambda ,\lambda ^{\prime }}Tr(e^{2i\pi \tau \stackrel{%
\sim }{L_{0}}}P_{\lambda ^{\prime }}\,A(0)\,P_{\lambda }\,A(\alpha )\,).
\label{eq1.7}
\end{equation}
In order to exhibit the functional equations for the chiral KMS states we
use the exchange algebra for the elementary field $A(\alpha )$ 
\begin{eqnarray}
(A)_{\delta \gamma }(\alpha )(A)_{\gamma \beta }(0) &=&\sum_{\gamma ^{\prime
}}\left[ R^{(\delta ,\beta )}\right] _{\gamma \gamma ^{\prime }}(A)_{\delta
\gamma ^{\prime }}(0)(A)_{\gamma ^{\prime }\beta }(\alpha ).  \nonumber \\
(A)_{\delta \gamma }(\alpha ) &\doteq &P_{\delta }A(\alpha )P_{\gamma }
\label{eq1.8}
\end{eqnarray}
to write (\ref{eq1.7}) as: 
\begin{equation}
{\cal F}_{\lambda \lambda ^{\prime }}(\alpha +\tau \mid \tau )=\sum_{\lambda
^{\prime \prime }}\left[ R^{(\lambda ^{\prime },\lambda ^{\prime })}\right]
_{\lambda \lambda ^{\prime \prime }}{\cal F}_{\lambda \lambda ^{\prime
\prime }}(\alpha \mid \tau )  \label{eq1.9}
\end{equation}
where the indices $\lambda ,\lambda ^{\prime },\lambda ^{\prime \prime }$
are labelling irreducible representations (sectors) of some chiral algebra
which satisfy certain fusion rules. The reader can find details in reference
\cite{RS}.

Next, we consider the action of the center of the conformal group, generated
by $N=\exp (2i\pi L_{0})$, on the primary field $A(\alpha )$ 
\begin{equation}
N\,P_{\lambda }\,A(\alpha )\,P_{\lambda ^{\prime }}\,N^{-1}=e^{2i\pi
(h_{\lambda ^{\prime }}-h_{\lambda })}P_{\lambda }\,A(\alpha )\,P_{\lambda
^{\prime }}.  \label{eq1.10}
\end{equation}
This provide us with another set of functional equations for the KMS states 
\begin{equation}
{\cal F}_{\lambda \lambda ^{\prime }}(\alpha +1\mid \tau )=e^{2i\pi
(h_{\lambda ^{\prime }}-h_{\lambda })}\,{\cal F}_{\lambda \lambda ^{\prime
}}(\alpha \mid \tau ).  \label{eq1.11}
\end{equation}
Therefore, knowing the exchange matrices $R$ and the conformal dimensions $%
h_{\lambda }$, we can (in principle) compute the KMS states by solving the
functional equations (\ref{eq1.9}) and (\ref{eq1.11}).

The critical Ising model is the simplest minimal conformal model with three
primary fields $\phi $, $\sigma $ and $\epsilon $ with conformal dimensions $%
h_\phi =0,\ h_\sigma =1/16$ and $h_\epsilon =1/2$ and central charge $c=1/2 $
.

The fusion rules for the field $\sigma $ are given by 
\[
\lbrack \sigma ][\phi ]=[\sigma ],\quad [\sigma ][\sigma ]=[\phi ]+[\epsilon
],\quad [\sigma ][\epsilon ]=[\sigma ] 
\]
and give a natural base to write the exchange $R$ matrices \cite{RS}: 
\[
R^{(\sigma ,\sigma )}=\left( 
\begin{array}{cc}
\lbrack R^{(\sigma ,\sigma )}]_{\phi \phi } & [R^{(\sigma ,\sigma )}]_{\phi
\epsilon } \\ 
\lbrack R^{(\sigma ,\sigma )}]_{\epsilon }\phi & [R^{(\sigma ,\sigma
)}]_{\epsilon \epsilon }
\end{array}
\right) =\left( 
\begin{array}{cc}
\frac{1}{2}(\eta +\eta ^{-3}) & \frac{1}{2}(\eta -\eta ^{-3}) \\ 
\frac{1}{2}(\eta -\eta ^{-3}) & \frac{1}{2}(\eta +\eta ^{-3})
\end{array}
\right) 
\]
\[
\lbrack R^{(\epsilon ,\epsilon )}]_{\sigma \sigma }=[R^{(\phi ,\phi
)}]_{\sigma \sigma }=\eta 
\]
where $\eta =\exp (-2i\pi h_{\sigma })=\exp (-i\pi /8)$.

Now we identify the elementary field $A(\alpha )$ with the primary field $%
\sigma $ and the equations (\ref{eq1.9}) and (\ref{eq1.11}) can be written
in a compact matricial form 
\begin{equation}
{\cal F}(\alpha +\tau \mid \tau )=M\ {\cal F}(\alpha \mid \tau ),\qquad 
{\cal F}(\alpha +1\mid \tau )=K\ {\cal F}(\alpha \mid \tau )  \label{eq1.12}
\end{equation}
where the chiral components ${\cal F}_{\lambda \lambda ^{\prime }}$ are
indexed as elements of the matrix column ${\cal F}$ and 
\begin{equation}
M=\left( 
\begin{array}{cccc}
0 & \alpha & \beta & 0 \\ 
\eta & 0 & 0 & 0 \\ 
0 & 0 & 0 & \eta \\ 
0 & \beta & \alpha & 0
\end{array}
\right) ,\ K=\limfunc{diag}\ (\eta ^{-1},\eta ,-\eta ,-\eta ^{-1})
\label{eq1.13}
\end{equation}
where 
\begin{equation}
\alpha =\frac{1}{2}(\eta +\eta ^{-3})\qquad \text{and\quad }\beta =\frac{1}{2%
}(\eta -\eta ^{-3})\text{ }  \label{eq1.14}
\end{equation}
Note that the non-vanishing entries of the matrix $K$ are eigenvalues of the
matrix $M$.

We now proceed to the solution of these functional equations. From (\ref
{eq1.9}) and (\ref{eq1.11}) we derive the following system of equations for
the chiral components 
\begin{eqnarray}
({\cal F}_{\sigma \phi }+{\cal F}_{\sigma \epsilon })(\alpha +1 &\mid &\tau
)=\eta ({\cal F}_{\sigma \phi }-{\cal F}_{\sigma \epsilon })(\alpha \mid
\tau )  \nonumber \\
({\cal F}_{\phi \sigma }+{\cal F}_{\epsilon \sigma })(\alpha +1 &\mid &\tau
)=\eta ^{-1}({\cal F}_{\phi \sigma }-{\cal F}_\epsilon \sigma )(\alpha \mid
\tau )  \nonumber \\
({\cal F}_{\sigma \phi }-{\cal F}_{\sigma \epsilon })(\alpha +1 &\mid &\tau
)=\eta ({\cal F}_{\sigma \phi }+{\cal F}_{\sigma \epsilon })(\alpha \mid
\tau )  \nonumber \\
({\cal F}_{\phi \sigma }-{\cal F}_{\epsilon \sigma })(\alpha +1 &\mid &\tau
)=\eta ^{-1}({\cal F}_{\phi \sigma }+{\cal F}_\epsilon \sigma )(\alpha \mid
\tau )  \label{eq1.15}
\end{eqnarray}
in the $1-$direction and 
\begin{eqnarray}
({\cal F}_{\sigma \phi }+{\cal F}_{\sigma \epsilon })(\alpha +\tau &\mid
&\tau )=\eta ({\cal F}_{\phi \sigma }+{\cal F}_{\epsilon \sigma })(\alpha
\mid \tau )  \nonumber \\
({\cal F}_{\phi \sigma }+{\cal F}_{\epsilon \sigma })(\alpha +\tau &\mid
&\tau )=\eta ({\cal F}_{\sigma \phi }+{\cal F}_{\sigma \epsilon })(\alpha
\mid \tau )  \nonumber \\
({\cal F}_{\sigma \phi }-{\cal F}_{\sigma \epsilon })(\alpha +\tau &\mid
&\tau )=\eta ({\cal F}_{\phi \sigma }-{\cal F}_{\epsilon \sigma })(\alpha
\mid \tau )  \nonumber \\
({\cal F}_{\phi \sigma }-{\cal F}_{\epsilon \sigma })(\alpha +\tau &\mid
&\tau )=\eta ^{-3}({\cal F}_{\sigma \phi }-{\cal F}_{\sigma \epsilon
})(\alpha \mid \tau )  \label{eq1.16}
\end{eqnarray}
in the $\tau -$direction.

Introducing the classical theta functions 
\begin{eqnarray*}
\Theta \left[ 
\begin{array}{c}
k_1 \\ 
k_2
\end{array}
\right] (\alpha &\mid &\tau )=\sum_{n=-\infty }^\infty \exp (i\pi
(n-k_1)^2\tau +2i\pi (n-k_1)(\alpha -k_2) \\
k_1,k_2 &\in &\{0,1/2\}
\end{eqnarray*}
to write the four Jacobi $\theta $-functions 
\[
\Theta _1=\Theta \left[ 
\begin{array}{c}
1/2 \\ 
1/2
\end{array}
\right] ,\ \Theta _2=\Theta \left[ 
\begin{array}{c}
1/2 \\ 
0
\end{array}
\right] ,\ \Theta _3=\Theta \left[ 
\begin{array}{c}
0 \\ 
0
\end{array}
\right] ,\ \Theta _4=\Theta \left[ 
\begin{array}{c}
0 \\ 
1/2
\end{array}
\right] 
\]
we have derived the following solution for the components of two-point KMS
function of the critical Ising model

\begin{equation}
\left( 
\begin{array}{c}
{\cal F}_{\phi \sigma } \\ 
{\cal F}_{\sigma \phi } \\ 
{\cal F}_{\sigma \epsilon } \\ 
{\cal F}_{\epsilon \sigma }
\end{array}
\right) =\frac{\left[ \frac{\Theta _1(\alpha \mid \tau )}{\Theta _1^{\prime
}(0\mid \tau )}\right] ^{-1/8}}{2\sqrt{\eta (\tau )}}\left( 
\begin{array}{c}
\lbrack \Theta _3(\frac \alpha 2\mid \tau )]^{1/2}+[\Theta _4(\frac \alpha
2\mid \tau )]^{1/2} \\ 
\lbrack \Theta _2(\frac \alpha 2\mid \tau )]^{1/2}+[-i\Theta _1(\frac \alpha
2\mid \tau )]^{1/2} \\ 
\lbrack \Theta _2(\frac \alpha 2\mid \tau )]^{1/2}-[-i\Theta _1(\frac \alpha
2\mid \tau )]^{1/2} \\ 
\lbrack \Theta _3(\frac \alpha 2\mid \tau )]^{1/2}-[\Theta _4(\frac \alpha
2\mid \tau )]^{1/2}
\end{array}
\right) .  \label{eq1.17}
\end{equation}
where $\eta (\tau )$ is the Dedekind $\eta $-function and $\Theta _1^{\prime
}(\alpha \mid \tau )=\frac \partial {\partial \alpha }\Theta _1(\alpha \mid
\tau )$.

Let us now examine the monodromy properties of ${\cal F}(\alpha \mid \tau )$
as $\alpha $ winds around a torus of periods $1$ and $\tau $. The monodromy
property of ${\cal F}(\alpha \mid \tau )$ as $\alpha \rightarrow \alpha +1$
is given by the action of the center of the conformal group (\ref{eq1.11}),
whereas the effect of $\alpha \rightarrow \alpha +\tau $ is given by the KMS
boundary condition (\ref{eq1.9}).

Instead of examining directly the modular properties of the components $%
{\cal F}_{\lambda \lambda ^{\prime }}$, we shall make use of the normalized
combinations of ${\cal F}$'s defined by 
\begin{equation}
{\cal F}_{\phi }={\cal F}_{\phi \sigma },\quad {\cal F}_{\sigma }=\frac{1}{%
\sqrt{2}}({\cal F}_{\sigma \phi }+{\cal F}_{\sigma \epsilon })\quad \text{%
and\quad }{\cal F}_{\epsilon }={\cal F}_{\epsilon \sigma }  \label{eq1.19}
\end{equation}
Another interesting limit is $\alpha \rightarrow 0$. It may be seen that in
this limit 
\begin{equation}
\lim_{\alpha \rightarrow 0}{\cal F}_{\lambda }(\alpha \mid \tau )\sim \chi
_{\lambda }(\tau )  \label{eq1.20}
\end{equation}
where $\chi _{\lambda }(\tau )$ are the characters of the critical Ising
model.

Consider now the effect of the modular transformations, $S:\tau \rightarrow
-\tau ^{-1}$ and $T:\tau \rightarrow \tau +1$ on the previously defined
combinations of the chiral KMS states. Using modular properties of the
Jacobi theta functions and the Poisson's formula it is easy to see that $S$
acts linearly on these combined states 
\[
{\cal F}_{\lambda }(\alpha \mid -\frac{1}{\tau })=\sum_{\lambda ^{\prime
}}S_{\lambda \lambda ^{\prime }}{\cal F}_{\lambda ^{\prime }}(\alpha \mid
\tau ), 
\]
whereas $T$ is diagonal 
\[
{\cal F}_{\lambda }(\alpha \mid \tau +1)=\sum_{\lambda }T_{\lambda \lambda }%
{\cal F}_{\lambda }(\alpha \mid \tau ). 
\]
Here $S$ is a symmetric unitary matrix , satisfying $S^{2}=(ST)^{3}.$ This
suffices to repeat Verlinde's steps \cite{Verlinde} and express the fusion
algebra matrix in terms of the matrix $S$.

With these expressions we have achieved our goal to identify the modular
properties of the chiral KMS states with the modular properties of
characters of the critical Ising model.

{\bf \flushleft  Acknowledgments}

It is ALS.'s pleasure to thank Profs. B. Schroer and K.-H. Rehren for useful
discussions.

\end{document}